\documentclass{article}

\usepackage[english]{babel}

\usepackage[letterpaper,top=2cm,bottom=2cm,left=3cm,right=3cm,marginparwidth=1.75cm]{geometry}

\usepackage{amsmath}
\usepackage{graphicx}
\usepackage[colorlinks=true, allcolors=blue]{hyperref}
\usepackage{subfig}

\title{Looking for the quantum aspects of gravity in the gravitational Aharonov-Bohm experiment}
\author{Ayda Najafzadeh}
\date{}
\begin{document}
\maketitle

\begin{abstract}
The detection of quantum aspects of gravity remains one of the most elusive challenges in modern physics. In this paper, we develop a comprehensive theoretical framework for the gravitational Aharonov-Bohm (AB) effect, extending previous classical models to a fully quantum description. By quantizing the gravitational field and modeling its interaction with atomic states, we derive a formulation for the gravitational AB phase mediated by gravitons. This framework uncovers key insights into the entanglement dynamics and coherence properties of quantum systems in weak gravitational fields. Our analysis suggests that the derived gravitational AB phase is consistent with classical predictions but reveals subtle quantum features, providing a robust basis for exploring the quantum nature of perturbative gravity. These findings offer a conceptual pathway for indirect detection of gravitons, enriching our understanding of gravity’s quantum underpinnings.
\end{abstract}

%\section{Introduction}
\section{Introduction}

 Bridging the gap between quantum mechanics and general relativity has been a fundamental question since the emergence of these two theories. It is believed that our experimental technologies cannot probe quantum gravitational phenomena directly due to the weakness of gravitational interactions at accessible scales. Gravitons—the hypothetical quantized carriers of the perturbative gravitational field—are central to the quest for quantum gravity. Their detection would provide concrete evidence that gravity itself exhibits quantum properties. Yet, the challenge of detecting gravitons directly has made this a long-standing problem, given the exceptionally weak gravitational coupling compared to other forces \cite{wallace2014deflating}\cite{dyson2013graviton}.

Recent developments in quantum experiments, particularly atom interferometry, have made it possible to observe gravitational effects with unprecedented precision \cite{bravo2023fluctuations}. One particularly promising avenue is the gravitational Aharonov-Bohm (AB) effect \cite{overstreet2022observation}, which draws on the successful observation of phase shifts caused by scalar electromagnetic potentials in quantum systems \cite{aharonov1959significance}. The gravitational analog of this effect, as recently demonstrated in experiments with Rubidium atoms inducing phase shifts in atom interferometers, provides an opportunity to explore quantum gravitational interactions at accessible scales. This work proposes an extension of these efforts by exploring the indirect detection of gravitons through a modified gravitational AB experiment.

Inspired by previous insights into the local generation of AB phases via quantum field interactions \cite{marletto2020aharonov}, we develop a framework where the presumed quantized particles of the gravitational field induces measurable phase shifts in a superposed atom interferometer. By extending classical descriptions to a fully quantum framework, we derive a formulation for the gravitational AB phase that incorporates quantized gravitational interactions. Our model predicts that an experimental setup could indirectly reveal the presence of gravitons through phase shift measurements, thus offering an indirect route to confirm the quantum nature of perturbative gravity.

The results obtained suggest that the gravitational AB phase derived here is consistent with classical predictions \cite{overstreet2022observation}, but reveals subtle effects consistent with quantum gravitational interactions. For example, by deriving the linear entropy in the gravitational AB experiment \cite{overstreet2022observation}, we concluded an entanglement measure 1,000 times stronger than the current experimental proposals for generation of entanglement between two superposed quantum masses \cite{marletto2017gravitationally}. This result is attributed to the classical source mass enhancing the interaction with the gravitational field. In fact, the structure of the gravitational AB experiment is similar to the Feynman thought experiment to use a single superposed quantum mass in a gravitational field \cite{feynman1957feynman}, which suggests a pathway toward the indirect detection of quantum features of gravity by exploiting the coherence features of the atomic states.

This paper is structured as follows: Section 2 provides an in-depth review of the electromagnetic and gravitational AB effects as theoretical foundations for this study. Section 3 outlines our fully quantum formalism for the gravitational AB effect, detailing the interaction dynamics of atomic states and gravitational fields. Section 4 presents proposed experimental configurations and methodologies for detecting the gravitational phase shifts attributable to gravitons, as well as potential experimental challenges and future directions, followed by our conclusions in Section 5.

%Future missions like the LISA spacecraft, which will utilize atom interferometry to detect gravitational waves in space \cite{folkner1998lisa}, offer a promising platform for conducting this experiment. Leveraging such technology could allow us to probe the quantum nature of gravity and detect gravitons indirectly at accessible scales.

%\href{https://www.overleaf.com/learn}{help library}, or head to our plans page to \href{https://www.overleaf.com/user/subscription/plans}{choose your plan}.

%\section*{Some examples to get started}
\section{The Aharonove-Bohm Effect}
\subsection{The Electromagnetic AB Effect}

\begin{figure}
  \centering
  \subfloat[The electromagnetic AB effect. Taken from \cite{griffiths2019introduction}]{\includegraphics[width=0.4\textwidth]{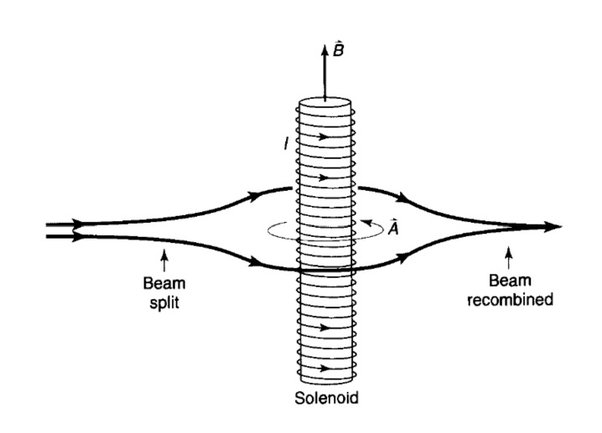}\label{fig:f1}}
  \hfill
  \subfloat[The scalar AB effect. Taken from \cite{kim2018electric}]{\includegraphics[width=0.3\textwidth]{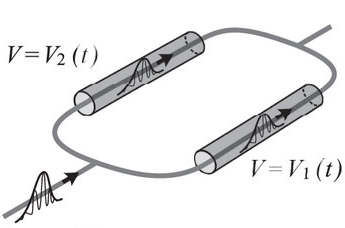}\label{fig:f2}}
  \caption{Aharonov-Bohm effect.}
\end{figure}

The Ehrenberg–Siday–Aharonov–Bohm (AB) effect \cite{aharonov1959significance,ehrenberg1949refractive} is a quantum phenomenon where a charged particle experiences a shift in its phase due to an electromagnetic potential, even though it moves through a region where the magnetic and electric fields are both zero. (Figure \ref{fig:f1}) 

The description of this phenomenon has been controversial over the last 7 decades. Traditionally, the AB effect has been understood as arising from the interaction between a particle and a classical electromagnetic vector potential \cite{aharonov1959significance}. The phase shift of the electron’s wave packet, which is superposed between the two arms of the interferometer, is known as the Aharonov-Bohm phase and is given by:\\
\begin{equation}\label{Eq:1}
{\Delta \phi}_{AB} = \frac{q}{\hbar}\oint_C \mathrm{\textbf{A}} \cdot  
 ~\mathrm{d} \mathrm{\textbf{l}},
\end{equation}

where, \textbf{A} is the vector potential, q is the charge, and C is the enclosed path of the superposed electron. This model reveals  that the potentials in quantum mechanics, particularly the vector potential, have observable effects even in the absence of the local field interactions \cite{aharonov2015comment,aharonov2016nonlocality}. 

More recent works, however, have reinterpreted the AB effect by focusing on local interactions, where the electron's influence on the source of the potential (e.g., a solenoid) produces the phase shift \cite{vaidman2012role}. This view suggests that, by considering the relative motion of particles, the AB effect can be explained through classical field interactions, without relying on potentials \cite{kang2017proposal}.

Additionally, quantum field theory offers a perspective where the phase shift arises from local quantum interactions mediated by entanglement between the charge and the photon field \cite{wallace2014deflating}. This fully quantum mechanical approach suggests that the AB phase, like other quantum phases, is generated locally through interactions between the particle’s quantum state and the surrounding electromagnetic field. Considering a charged particle of mass 
m, charge q, and momentum p, superposed around a solenoid with current density j, the local quantum AB phase is calculated to be \cite{marletto2020aharonov}: \\
\begin{equation}
{\Delta \phi}_{AB} \left(\mathbf{\textbf{r}}_{\mathbf{L}}, \mathbf{\textbf{r}}_{\mathrm{R}}\right)= \frac{2tq}{m \hbar \epsilon_0 c^2} \int_V \frac{\mathrm{\textbf{p}} \cdot \mathrm{\textbf{j}}\left(\mathrm{\textbf{x}}-\mathrm{\textbf{r}}_{\mathrm{s}}\right)}{\left|\mathrm{\textbf{r}}_{\mathrm{c}}-\mathrm{\textbf{x}}\right|} 
 ~\mathrm{d}^3 \mathrm{\textbf{x}},
\end{equation}

where, $\textbf{r}_s$ and $\textbf{r}_c$ represent the positions of the solenoid and the charged particle at time t, and V is the standard quantization volume for the photon field. Interestingly, this local AB phase is related to the positions of the charge and the solenoid at each moment during the trajectory of the electron.  Two experimental proposals have been suggested for detecting this phase shift \cite{marletto2020aharonov}. One approach involves local tomography to track the AB phase over time, while the second proposes demonstrating the role of photons in generating the AB phase. In the latter case, the interference loop must be closed before any photon exchange occurs. We propose that this latter approach could serve as the key to the indirect detection of gravitons. 

Building on the electromagnetic case, the gravitational Aharonov-Bohm effect offers a similar framework for understanding quantum gravitational interactions, where gravitational fields induce phase shifts in quantum systems. In the following section, we explore how this effect could provide an indirect method to detect gravitons. 

\subsection{The Gravitational AB Effect}

%First you have to upload the image file from your computer using the upload link in the file-tree menu. Then use the includegraphics command to include it in your document. Use the figure environment and the caption command to add a number and a caption to your figure. See the code for Figure \ref{fig:frog} in this section for an example.
The recently observed gravitational Aharonov-Bohm (AB) effect \cite{overstreet2022observation} is analogous to the scalar AB effect, where the electric potential difference between the two paths of a superposed particle results in phase shifts that create an interference pattern (Figure \ref{fig:f2}).  In this case, however, the gravitational potential takes the place of the electric potential, influencing the particle’s trajectory and resulting in observable quantum interference.
%Note that your figure will automatically be placed in the most appropriate place for it, given the surrounding text and taking into account other figures or tables that may be close by. You can find out more about adding images to your documents in this help article on \href{https://www.overleaf.com/learn/how-to/Including_images_on_Overleaf}{including images on Overleaf}.
%\subsection{How to add Tables}

%Use the table and tabular environments for basic tables --- see Table~\ref{tab:widgets}, for example. For more information, please see this help article on \href{https://www.overleaf.com/learn/latex/tables}{tables}. 

\begin{figure}
\centering
\includegraphics[width=0.48\linewidth]{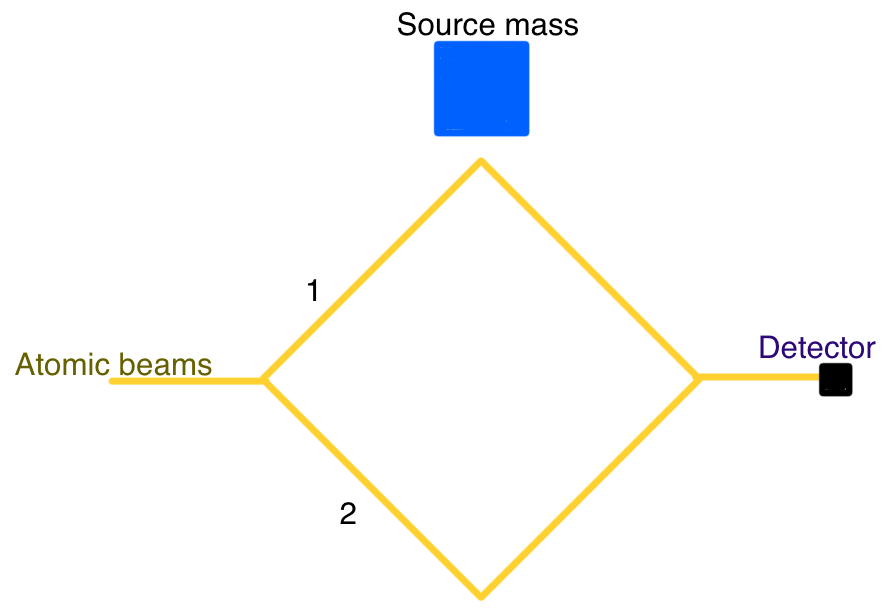}
\caption{\label{fig:Sourc mass}Gravitational AB experiment}
\end{figure}

%\begin{table}
%\centering
%\begin{tabular}{l|r}
%Item & Quantity \\\hline
%Widgets & 42 \\
%Gadgets & 13
%\end{tabular}
%\caption{\label{tab:widgets}An example table.}
%\end{table}
In the gravitational AB experiment, an atomic interferometer setup is employed to detect the influence of a source mass on the phase shifts of each arm of the interferometer. Focusing only on the effect of the gravitational potential due to the source mass, and suppress other experimental contributions to observe this effect, one can show the simplified figure of this experiment as Figure \ref{fig:Sourc mass}. According to the Feynman path integral, the classical action along the trajectories of the atoms generates different phase shifts for each arm. For instance, the phase difference between the trajectories of an atom interferometer is given by \cite{overstreet2022observation}:
\\
\begin{equation}\label{Eq:3}
\phi_{\Delta S}= \frac{\Delta S}{\hbar}=\frac{m}{\hbar} \int\left(\left[V\left(x_1, t\right)-V\left(x_2, t\right)\right]-\frac{\Delta x}{2}\left[\frac{\partial V\left(x_1, t\right)}{\partial x}+\frac{\partial V\left(x_2, t\right)}{\partial x}\right]\right) d t \quad.
\end{equation}
\\
Here, V represents the gravitational potential of the source mass on each arm of the interferometer with atomic mass of m, and $\Delta x$ is the separation distance between the two interferometer arms. The derivative terms in this integral relate to the kinetic energy of atoms, which simplified using other terms in the interferometry experiment \cite{overstreet2021physically}. The phase shift due to the gravitational interaction between the atoms and the source mass is the desired Aharonov-Bohm phase, which can be experimentally observed when the source mass is positioned at an optimal distance to minimize other phase contributions, including the derivative term in equation (\ref{Eq:3}):
\\
\begin{equation}
{\Delta \phi}_{AB}=\phi_{\Delta S}=\frac{m}{\hbar} \int\left[V\left(x_1, t\right)-V\left(x_2, t\right)
%-V\left(x_3, t\right)+V\left(x_4, t\right)
\right] d t \quad.
\end{equation}
\\
While it has been shown that using the semi-classical model to describe the phase shifts is more efficient compared with the Mid-point theorem \cite{overstreet2022observation}, implementing a quantized gravitational field brings new insights into the gravitational interactions at quantum level where the interactions between masses are mediated by gravitons. This leads us to the next section, where we explore a fully quantum description of the gravitational AB effect.

\section{Fully Quantum Description of AB Phase}

While the observed gravitational Aharonov-Bohm effect can be an indirect evidence of the quantum nature of gravity \cite{overstreet2021physically}\cite{feng2023conservation}, the classical model of the gravitaional field might falls short of capturing the full quantum nature of gravitational interactions. To investigate these interactions more deeply, we turn to a fully quantum mechanical framework, where the weak gravitational field is quantized, and the interactions between masses are mediated by gravitons.

\begin{figure}
\centering
\includegraphics[width=0.5\linewidth]{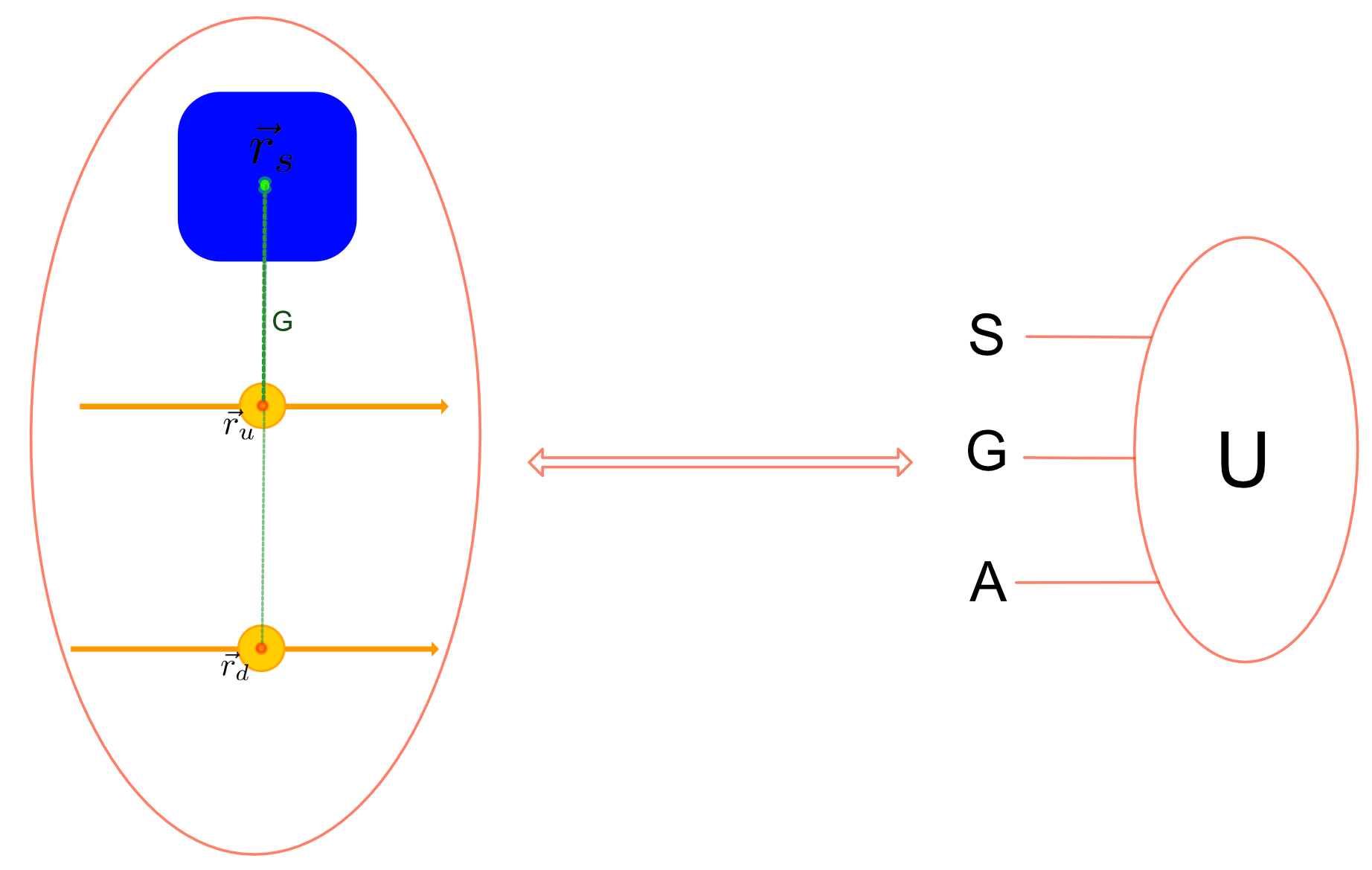}
\caption{\label{fig:Network}The quantum network of the gravitational interaction between masses}
\end{figure}

In this quantum framework, we model the interaction between the atom, the source mass, and the gravitational field using a quantum network (figure \ref{fig:Network}). In this representation, the atom (A), the graviton (G), and the source mass (S) form a network of interactions, with each element contributing to the generation of the Aharonov-Bohm phase. The interaction between these components is described through a U gate, governing the quantum evolution of the system.

Consider the atom located at $\textbf{r}_c$, the source mass at $\textbf{r}_s$ and the gravitational field F, whose observables act on the space $\mathcal{H} = \mathcal{H}_m \otimes \mathcal{F}_G \otimes \mathcal{H}_M$, where $ \mathcal{H}_m$ is the Hilbert space of atomic quantum state, $\mathcal{H}_M$ is the Hilbert space of the source mass, and the $\mathcal{F}_G$ denotes the Fock space of the gravitons' field. Now the initial state of the system before action of the gate is:
\\
\begin{equation}
|\psi(0)\rangle=\frac{1}{\sqrt{2}}\left(\left|m_d\right\rangle \otimes|M\rangle \otimes|0\rangle_g+\left|m_u\right\rangle \otimes|M\rangle \otimes|0\rangle_g\right) \quad.
\end{equation}\\ 
Here, $\left|0\right\rangle_g$ is the vacuum state of the gravitational field, $\left|m\right\rangle_d$ and $\left|m\right\rangle_u$ are the states of the atom in the lower and upper arms of the interferometer, while $\left|M\right\rangle$ is the state of the source mass. These mass states are the localized states of a neutral scalar field's excitation that are given by the creation and annihilation operators of their fields \cite{bose2017spin}. 

The total Hamiltonian for the system accounts for the free energy of the masses and the gravitational field, as well as their interactions. The first terms represent the free energy of the masses, while the interaction terms describe how the gravitational field (mediated by gravitons) couples to the masses:
\begin{equation}\label{Eq:Hamiltonian}
\begin{aligned}
    H &= mc^2 \left(a_d^\dagger a_d + a_u^\dagger a_u\right)
    + Mc^2 a_s^\dagger a_s 
    + \sum_{k, \lambda} \hbar \omega_k b_{k,\lambda}^\dagger b_{k,\lambda} \\
    &\quad - \frac{1}{\hbar} \sum_{k, \lambda} c_\lambda \sqrt{\frac{2\pi G}{\hbar \omega_k V}}
    \Bigg[
    ma_d^\dagger a_d \left(b_{k,\lambda} e^{i \mathbf{k} \cdot \mathbf{r}_d} 
    + b_{k,\lambda}^\dagger e^{-i \mathbf{k} \cdot \mathbf{r}_d}\right) \\
    &\quad \,\, + ma_u^\dagger a_u \left(b_{k,\lambda} e^{i \mathbf{k} \cdot \mathbf{r}_u} 
    + b_{k,\lambda}^\dagger e^{-i \mathbf{k} \cdot \mathbf{r}_u}\right) \\
    &\quad \,\, + Ma_s^\dagger a_s \left(b_{k,\lambda} e^{i \mathbf{k} \cdot \mathbf{r}_s} 
    + b_{k,\lambda}^\dagger e^{-i \mathbf{k} \cdot \mathbf{r}_s}\right)
    \Bigg].
\end{aligned}
\end{equation}
Here, ${a}_\xi^{\dagger}$ stands for the creation operator of a mass $m_\xi$ in a spatially localized wavepacket around the point $\textbf{r}_\xi$, and $b_{k, \lambda}^{\dagger}$ is the bosonic creation operator of the gravitational field in mode $\textbf{k},\lambda$. In this equation, the coupling constant of the gravitational interaction between the field and an arbitrary mass $m_l$ is given by:
\\
\begin{equation}
 g_l = m_l c \sqrt{\frac{2 \pi G }{\hbar \omega_k V}},
\end{equation}
 \\
where, V is the standard quantisation volume, G is the gravitational constant, and $\omega_k$ represents the graviton frequency of the \textbf{k}-th mode.

The time-dependent state of the system evolves according to this Hamiltonian:
\\
\begin{equation}
\begin{aligned}
  |\psi(t)\rangle=e^{-i H t / \hbar}|\psi(0)\rangle \quad. \end{aligned}
\end{equation}    
\\
The solution of this dynamical evolution is derived as \cite{bose1997preparation}\cite{bose2017spin}:
\\
\begin{equation}\label{Eq:State}
  \begin{aligned}
 |\psi(t)\rangle =\frac{1}{2} \sum_{\xi \in\{u, d\}} e^{i \sum_{\textbf{k}, \lambda} \frac{t}{\omega_k}\left|g_s e^{i \textbf{k} \cdot \textbf{r}_s}+g_\xi e^{i \textbf{k} \cdot \textbf{r}_\xi}\right|^2}\left|m_{\xi}\right\rangle|M\rangle \prod_{\textbf{k},\lambda}|\frac{1-e^{-i\omega_k t}}{\omega_k }\left(g_s e^{i\textbf{k} \cdot \textbf{r}_s}+g_\xi e^{-i \textbf{k} \cdot \textbf{r}_\xi}\right)\rangle,
\end{aligned}
\end{equation} 
where, $g_s$ and $g_\xi$ are the coupling constants of the source mass and the atomic arms respectively. The calculated state has two main terms related to the interaction of each arm $(\xi \in\{u, d\})$ of the atomic interferometer with the gravitational field. In fact, each arm being entangled with the gravitons' field receives a phase shift related to their interaction. After the interaction, the reduced coherent states of the gravitational field and the atomic state are clearly mixed states of the following forms:
\\
\begin{equation}
\rho_{\alpha}= \frac{1}{2} \sum_{\xi \in\{u, d\}}|\alpha_{\xi}\rangle \langle\alpha_{\xi}|,
\end{equation}
\\
\begin{equation}
\rho_{m}= \frac{1}{2} \left(|m_u\rangle\langle m_u\right|+|m_d\rangle\langle m_d|+ e^{i\Delta\phi_{AB}}\langle \alpha_d|\alpha_u\rangle \left|m_u\rangle\langle m_d\right|+ e^{-i\Delta\phi_{AB}}\langle \alpha_u|\alpha_d\rangle \left|m_d\rangle\langle m_u| \right),
\end{equation}
\\
in which, $\left|\alpha_{\xi}\right\rangle$ is the coherent state of the gravitational field after the interaction at $\textbf{r}_{\xi}$. This information masking is an evident of the entanglement between gravitational field and the atoms. As a result, we can use the  Von Neumann  entropy to have a sense of this entanglement. In this scenario, it is evident that the gravitational interaction is weak. Therefore, we can use the linear entropy to describe the entanglement of each arms to the gravitational field:
\\
\begin{equation}\label{Eq:linear enthropy}
\begin{aligned}
& S_L=1-\operatorname{Tr}\left(\rho_\alpha^2\right) =\frac{1}{2} \left(1-\left|\left\langle\alpha_d \mid \alpha_u\right\rangle\right|^2\right)\approx\frac{\left|\alpha_d-\alpha_u\right|^2}{2} \approx 10^4 \frac{m^2}{m_p^2} \quad,
\end{aligned}
\end{equation}
\\
where, $m_p$ is the Planck mass, and m is the mass of atom. The derivation of this equation can be found in appendix A. This entropy can be considered as a counterpart of the one which is derived in the GIE experiment \cite{marletto2017gravitationally}, except, it is $10^4$ times stronger than the one derived for those kinds of experiments, as it is expected by an amplified gravitational field using a massive object \cite{pedernales2022enhancing}. Consequently, it might be easier to detect this entanglement experimentally. In fact, this experimental configuration is similar to the Feynman thought experiment on the detection of quantum nature of gravity using a quantum superposed mass in a gravitational field \cite{feynman1957feynman}. Putting the masses of the atom interferometery experiment \cite{overstreet2022observation}, it can be seen that this entanglement is in order of $10^{-29}$. 
%\textit{It is easy to show that this number is proportional to the coherent terms of the reduced density matrix of the atomic states. It shows how this interaction reduces the coherency between the atomic beams, and prevent us to observe the interfernce pattern. Despite the fact that this iterfernce pattern might contains information about the quantumness of gravity, it is believed that the false loss of quantum coherency \cite{leggett2002qubits}\cite{unruh2000false}, cannot be distinguished from a slowed classical de-phasing process \cite{marletto2024quantum}, in which in both cases, we are able to observe the quantum interference patterns after the gravitational interactions.}
Interestingly, this equation does not depend on the mass of the source mass. It simply shows that the effect of the source mass is mapped into the coherent state of the gravitational field. While this interaction produces entanglement between the atoms and the gravitational field, we expect that it has no effect on the state of the classical source mass.

Finally, the Aharonov-Bohm phase can be derived from the phase terms in this quantum evolution by integrating over the continuum of the graviton field's wave-number and subtracting the phase shifts between the two arms of the interferometer. The resulting phase difference, ${\Delta \phi}_{AB}$, is given by:
\begin{equation}
\begin{aligned}
 {\Delta \phi}_{AB}=\varphi_u-\varphi_d&=\operatorname{Re}\left\{V t \int d^3 \textbf{k} \frac{4 \pi G M m_u}{\hbar|\textbf{k}|^2 V} e^{i \textbf{k} \cdot\left( \textbf{r}_s - \textbf{r}_u\right)}\right\}-\operatorname{Re}\left\{V t \int d^3 \textbf{k} \frac{4 \pi G M m_d}{\hbar|\textbf{k}|^2 V} e^{i \textbf{k} \cdot\left( \textbf{r}_s - \textbf{r}_d\right)}\right\} \\
 &=\frac{G M t}{\hbar}\left(\frac{m}{\left|\textbf{r}_u-\textbf{r}_s\right|}-\frac{m}{\left|\textbf{r}_d-\textbf{r}_s\right|}\right) \quad. \\
\end{aligned}
\end{equation}
\\
Interestingly, this derived gravitational AB phase is similar to its classical counterpart. In fact, we showed that the effect of a quantised gravitational field cannot be observed directly in the phase shift of the interferometer. According to the fact that we used the linearized Einstein equation, it is evident that this phase shift is related to the Newtonian gravitational potential. 

By quantizing the gravitational field and considering the interactions between particles as mediated by gravitons, we gain deeper insights into the quantum nature of gravity. This quantum formalism provides a framework for testing gravitational interactions at the quantum level, paving the way for the indirect detection of gravitons in future experiments.
%Comments can be added to your project by highlighting some text and clicking_s = g ``Add comment'' in the top right of the editor pane. To view existing comments, click on the Review menu in the toolbar above. To reply to a comment, click on the Reply button in the lower right corner of the comment. You can close the Review pane by clicking its name on the toolbar when you're done reviewing for the time being.

\section{Experimental proposals}
%atom lens cooling to cool the atoms in spacecraft

To indirectly demonstrate the existence of quantum particles such as gravitons, we build on methods used in electromagnetic Aharonov-Bohm experiments \cite{marletto2020aharonov}. In the gravitational case, we propose two experimental configurations to detect gravitons by observing their influence on the phase shift of an interferometer’s arm. Both proposals rely on quantum entanglement and phase coherence, but differ in the timing and interaction between the interferometer and the gravitons.

\textbf{One-Arm Entanglement:} In the first configuration, one arm of the interferometer interacts with gravitons while the other arm remains unaffected. This results in a detectable phase difference compared to the gravitational AB phase, since only one arm of the interferometer entangles with the gravitons. The interferometer loop must be closed before graviton exchange occurs with the second arm, ensuring that only one arm interacts with the quantum gravitational field. This setup requires careful control of distances, as the arm interacting with the gravitational source must be much closer to the source mass than the second arm.

By maintaining a significant distance between the two interferometer arms and the source, the gravitational interaction generates a measurable phase shift in one arm before gravitons reach the other arm. This selective entanglement modulates the gravitational AB phase, resulting in a distinct phase shift signature that differs from the standard AB effect. 

\textbf{No-Arm Entanglement:} In the second configuration, the interferometer loop is closed before either arm can interact with the gravitons. This ensures that no entanglement between the interferometer and the gravitational field occurs before the loop closure. Under these conditions, the experiment seeks to detect whether gravitons still induce a phase shift in the interferometer despite no direct entanglement with the arms. If no phase shift is observed, this would suggest that graviton-mediated interactions are necessary for generating the gravitational AB phase, indirectly confirming the existence of gravitons.

This setup is particularly challenging, as it requires measuring low-frequency gravitational fields with extreme precision. The absence of entanglement and phase shift would provide strong evidence for the role of gravitons in gravitational interactions, further solidifying the quantum nature of gravity.

\subsection*{\small Challenges and Feasibility}
Both experimental configurations face challenges in mitigating gravitational noise, particularly from Earth’s gravitational gradient. However, the use of atom interferometers offers a promising solution, as they are highly sensitive to small gravitational perturbations. Atom interferometers have already been used to detect gravitational effects with unprecedented precision, making them an ideal candidate for these experiments.

The primary challenge lies in maintaining the necessary distance between the interferometer arms and the source mass, and controlling the timing of the loop closure to ensure that graviton interactions occur (or do not occur) as intended. By measuring small changes resulting from differences in relative entanglement between the subsystems, we can succeed in detecting gravitons. This requires advanced precision measurement tools and careful calibration of the experimental setup.

\subsection*{\small Mitigating Gravitational Interferences with the LISA Project}
The LISA project (Laser Interferometer Space Antenna), a space-based gravitational wave detector scheduled for deployment in the 2030s, provides an ideal platform for this experimental proposal. LISA will use laser interferometry to detect gravitational waves at cosmological scales, operating in space to avoid the noise and disturbances caused by Earth’s gravitational fields.

LISA’s ability to detect gravitational waves with high precision makes it a powerful tool for exploring quantum aspects of gravity. Our proposal to use atom interferometers aboard LISA could extend the mission’s scope beyond classical wave detection, enabling us to investigate whether gravitons mediate gravitational interactions at the quantum level. This integration of quantum concepts into LISA’s framework could open new avenues for indirect graviton detection.

\subsection*{\small Comparison to Existing Experiments}
Existing gravitational wave detectors, such as LIGO and Virgo, have demonstrated impressive sensitivity in detecting spacetime distortions caused by classical gravitational waves. These detectors rely on measuring the strain induced by passing gravitational waves on interferometer arms that are kilometers in length. While these setups are highly effective for detecting large-scale classical gravitational waves, they are not designed to probe quantum aspects of gravity, such as individual graviton interactions.

Our proposed experiment, in contrast, seeks to explore the quantum nature of gravity by focusing on phase shifts induced by graviton interactions at much smaller scales. Unlike LIGO and Virgo, which measure spacetime distortions over macroscopic distances, our approach leverages atom interferometry to detect minute phase shifts resulting from quantum gravitational effects. 

Our approach, by focusing on phase coherence and the Aharonov-Bohm effect, offers a complementary method for probing quantum gravitational effects. By observing the absence or presence of a phase shift in the interferometer arms, we can indirectly infer the role of gravitons in gravitational interactions. This method circumvents some of the challenges associated with detecting entanglement directly, providing a more accessible route to exploring the quantum nature of gravity in controlled experimental conditions.

In addition, proposals such as graviton detection using quantum sensors \cite{tobar2024detecting}, rely on measuring mechanical systems' interactions with gravitons. While these methods seek to observe graviton-mediated energy transitions, our proposal emphasizes phase detection through interferometry, offering an approach that is distinct in its focus on phase shifts and quantum coherence. This makes our experiment highly sensitive to the graviton’s influence on superposed quantum systems, positioning it as a complementary method in the broader effort to detect quantum gravitational effects.

%To indirectly prove the existence of mediating quantum particles like gravitons, one can use the same idea in the electromagnetic AB experiment\cite{marletto2020aharonov}. If the interferometer loop is closed before the exchange of any mediating particle, this formalism predicts that no AB phase would be observed. This simple proposal might be acquired by the gravitational fields of massive objects in far distances. The only difficulty is to use several atom interferometers in order to mitigate the effects of other gravitational sources, like the earth's gradient gravity in the gravitational AB experiment\cite{overstreet2022observation}. In fact, using other atom interferometers enable us to implement metric perturbations in background Minkowski spacetime. As a result, the effect of a source mass can be modeled as a perturbation in this background. Gravitons are the excitation modes of this perturbed field. The Lisa project is one the candidates for doing this experiment in the future.   

\section{Conclusion}

Observation of evidences for quantumness of gravity highlight the importance of more experimental proposals to measure the characteristic of gravity \cite{feng2023conservation}. Gravitons are the most straightforward theoretical quantum aspects of gravity. However, it is important to notice the difference between gravitons and quantum gravity. The former are the components of weak gravitational perturbations, while the later refers to the basic building blocks of the spacetime \cite{hsiang2024graviton}. Moreover, quantum gravity is related to the Planck length scales whic is inaccessible with current technologies. On the other hand, "gravitons can exist at all length scales greater than the Planck length because once a spacetime manifold exists, one can consider weak perturbations on this classical entity and quantize them \cite{hsiang2024graviton}." So, what does the indirect detection of gravitons show us? The answer is that it confirms the quantum nature of perturbative  gravity. "Then, these gravitons can be seen as a low energy effective field theory of the basic constituents of the spacetime, or as the quantized collective excitation modes of spacetime \cite{hu2020semiclassical}\cite{donoghue1994general}."

\section*{Acknowledgments}
I am deeply grateful to Professors Vahid Karimiopur, Vlatko Vedral, Bahram Mashoon and Nima Khosravi for their invaluable guidance and support throughout this project. Their insights and expertise have been instrumental in shaping this work, and their encouragement has been a source of inspiration.

%\LaTeX{} is great at typesetting mathematics. Let $X_1, X_2, \ldots, X_n$ be a sequence of independent and identically distributed random variables with $\text{E}[X_i] = \mu$ and $\text{Var}[X_i] = \sigma^2 < \infty$, and let

\bibliographystyle{unsrt}
\bibliography{output}
%\section*{Appendix A: Calculation of The Post-interaction State}
%%%%%%%%%%%%%%%%%%%%%%%%%%%%%%
\newpage
\section*{Appendix A: Calculation of $|\psi(t)\rangle$}

This derivation follows the framework presented in \cite{bose1997preparation}, with modifications tailored to the specific dynamics of this system. The Hamiltonian in Equation (\ref{Eq:Hamiltonian}) is the starting point, and we simplify it by excluding the full modes of the gravitational field:
\begin{equation}
\begin{aligned}
    H = &\hbar \omega b^{\dagger} b+\hbar \omega^{\prime}\left(a_u^{\dagger} a_u + a_d^{\dagger} a_d\right)+\hbar \omega^{\prime \prime} a_s^{\dagger} a_s\\
    & \left. -\hbar\left[g a_u^{\dagger} a_u\left(b e^{i \textbf{k}\cdot\textbf{r}_u}+b^{\dagger} e^{-i \textbf{k} \cdot \textbf{r}_u}\right)\right.+ g a_d^{\dagger} a_d\left(b^{\dagger} e^{i \textbf{k} \cdot \textbf{r}_d}+b^{\dagger} e^{-i \textbf{k} \cdot \textbf{r}_d}\right)+g_s a_s^{\dagger} a_s\left(b e^{i \textbf{k} \cdot \textbf{r}_s}+b e^{-i \textbf{k} \cdot \textbf{r}_s}\right)\right],
\end{aligned}
\end{equation}

Where,\\
\begin{equation}
\hbar\omega^{\prime} = mc^2,  \quad\quad \hbar\omega^{\prime\prime} = Mc^2.    
\end{equation}

Then, we rewrite the Hamiltonian in a simplified form. 

\begin{equation}
    H=\hbar \omega\left[N_{\text {tot }}+b^{\dagger} b-b A-b^{\dagger} A^{\dagger}\right], 
\end{equation}
Where, 
\begin{equation}
     A=\frac{g}{\omega} a_d^{\dagger} a_d e^{i \textbf{k} \cdot \textbf{r}_d}+\frac{g}{\omega} a_u^{\dagger} a_u e^{i \textbf{k} \cdot \textbf{r}_u}+\frac{g_s}{\omega} a_s^{\dagger} a_s e^{-i \textbf{k} \cdot \textbf{r}_s}, \quad N_{\text {tot }}=\frac{\omega^{\prime}}{\omega}\left(a_u^{\dagger} a_u^{\dagger}+ a_d^{\dagger} a_d\right)+\frac{\omega^{\prime \prime}}{\omega} a_s^{\dagger} a_s .\\
\end{equation}
Now, we might define a displacement operator.
\begin{equation}
   T=e^{\left(b A-b^{\dagger} A^{\dagger}\right)},
\end{equation}
In which, one can use of the Baker–Campbell–Hausdorff formula to derive:
\begin{equation}
    T b T^{\dagger}=b+A^{\dagger}, \quad  T b^{\dagger} T^{\dagger}=b^{\dagger}+A.
\end{equation}
Therefore, we can rewrite the time-evolution operation by use of this displacement operator as:
\begin{equation}
    U(t)= T^{\dagger}T U T^{\dagger}T= T^{\dagger}\left(e^{-i \omega t N_{\text {bot }}}\right)\left(e^{i \omega t A A^{\dagger}}\right)\left(e^{-i \omega t b^{\dagger} b}\right)T =\left(e^{-i \omega t N_{tot}}\right)\left(e^{i \omega t A^{\dagger} A}\right) e^{-\left(b A-b^{\dagger} A^{\dagger}\right)} e^{-i \omega t b^{\dagger} b} e^{\left(b A-b^{\dagger} A^{\dagger}\right)}. \\
\end{equation}
Then, by use of the Baker–Campbell–Hausdorff formula one can reach to the helpful expression of the $U(t)$:
\begin{equation}
    U(t)= e^{-i \omega t N_{tot}} e^{\left(i A^{\dagger} A (\omega t-\sin (\omega t)\right)}e^{\left(b^{\dagger} A^{\dagger}\left(1-e^{-i \omega t}\right)-b A\left(1-e^{i \omega t}\right)\right)}e^{-i \omega t b^{\dagger} b}.
\end{equation}
Then, by applying this unitary on the initial state of the system and ignoring the global phase we have:
\begin{equation}
    |\psi(t)\rangle =\frac{1}{2} \sum_{\xi \in\{u, d\}} e^{i\frac{t}{\omega_k}\left|g_s e^{i \textbf{k} \cdot \textbf{r}_s}+g_\xi e^{i \textbf{k} \cdot \textbf{r}_\xi}\right|^2}\left|m_{\xi}\right\rangle|M\rangle |\frac{1-e^{-i\omega_k t}}{\omega_k }\left(g_s e^{i_\textbf{k} \cdot \textbf{r}_s}+g_\xi e^{-i \textbf{k} \cdot \textbf{r}_\xi}\right)\rangle
\end{equation}
The final state, Equation (\ref{Eq:State}), can be obtained by considering the full mode structure of the gravitational field.

%%%%%%%%%%%%%%%%%%%%%%%%%%%%%%%%%%%%%%%%%%%%%%

\section*{Appendix B: Calculation of $S_L$}

As we calculated, the time dependent state of the system can be written as a simple form as:
\\
\begin{equation}
\begin{aligned}
& |\psi(t)\rangle=\frac{1}{\sqrt{2}}\left(c_u\left|m_u\right\rangle\left|\alpha_u\right\rangle+c_d\left|m_d\right\rangle\left|\alpha_d\right\rangle\right)|M\rangle \quad.
\end{aligned}
\end{equation}
\\
So, the density matrix would be given by:
\\
\begin{equation}
\begin{aligned}
 \rho =|\psi \rangle \langle \psi|=\frac{1}{2}|M \rangle \langle M| \otimes&\left(\left|c_u\right|^2\left|m_u \rangle \langle m_u\right| \otimes \left|\alpha_u \rangle \langle \alpha_u\right|+\left|c_d\right|^2\left|m_d \rangle \langle m_d\right| \otimes \left|\alpha_d \rangle \langle \alpha_d\right|\right.\\
& \left.+ c_u c_d^*\left|m_u \rangle \langle m_d\right| \otimes\left|\alpha_u \rangle \langle \alpha_d\right|+c_u^* c_d\left|m_d \rangle \langle m_u\right| \otimes\left|\alpha_d \rangle \langle \alpha_u\right|\right) \quad.\\
\end{aligned}
\end{equation}
\\
Now, we can trace out the state of masses to reach the state of the gravitational field $\rho_{\alpha}$:
\\
\begin{equation}
\begin{aligned}
& \rho_\alpha=\operatorname{tr}_{m_\xi, M}(\rho)=\frac{1}{2 }\left(\left|c_u\right|^2\left|\alpha_u \rangle \langle \alpha_u\right|+\left|c_\alpha\right|^2\left|\alpha_\alpha \rangle \langle \alpha_d\right|\right), \\
\end{aligned}
\end{equation}
\\
where, $c_u$ and $c_d$ are the phases, which their norm is equal to 1. To derive the linear enthorpy $S_L=1-\operatorname{Tr}\left(\rho_\alpha^2\right)$, we need the to calculate:
\\
\begin{equation}
\begin{aligned}
& \rho_\alpha^2=\frac{1}{4} \left(|\alpha_u \rangle \langle \alpha_u\left|+\left|\alpha_d \rangle \langle \alpha_d\right|+\left\langle\alpha_u \mid \alpha_d\right\rangle\right| \alpha_u \rangle \langle\alpha_d|+\left\langle\alpha_d \mid \alpha_u\right\rangle\left|\alpha_d \rangle \langle \alpha_u\right|\right) \quad.
\end{aligned}
\end{equation}
\\
So, we have:
\begin{equation}\label{trace}
\begin{aligned}
& \operatorname{tr}\left(\rho_\alpha^2\right)=\frac{1}{2}\left|\left\langle\alpha_u \mid \alpha_d\right\rangle\right|^2+\frac{1}{2}
%\frac{1}{4} \int\left(\left|\left\langle\alpha \mid \alpha_u\right\rangle\right|^2+\left|\left\langle\alpha \mid \alpha_d\right\rangle\right|^2\right) \frac{d^2 \alpha}{\pi} \quad.\\
\end{aligned}
\end{equation}
Here, we used the completeness relation between the coherent states:
$$
\begin{aligned}
 \int |\alpha\rangle\langle\alpha| \frac{d^2 \alpha}{\pi} = 1,
\end{aligned}
$$
where,
$$
\begin{aligned}
\alpha=r e^{i \theta} , \quad d \alpha=r d r d \theta \quad.\\
\end{aligned}
$$
Now, using the equation (\ref{Eq:linear enthropy}), we can derive:
\\
\begin{equation}
\begin{aligned}
& \left|\left\langle\alpha_d \mid \alpha_u\right\rangle\right|^2=e^{-\left|\alpha_d-\alpha_u\right|^2}
%=e^{-\left(r^2+\textbf{r}_u^2-2 r \textbf{r}_u \cos \left(\theta-\theta_u\right)\right)} \quad.\\
\end{aligned}
\end{equation}
%Putting this into the integral, we have:
%\begin{equation}
%\begin{aligned}
%& \int\left|\left\langle\alpha \mid \alpha_u\right\rangle\right|^2\frac{d^2 \alpha}{\pi} =\frac{1}{\pi} \int_0^{\infty} \int_0^{2 \pi} e^{-r^2-\textbf{r}_u^2+2 r \textbf{r}_u \cos \left(\theta-\theta_u\right)} r d r d \theta=2e^{{-\textbf{r}_u}^2} \int_0^{\infty} e^{-r^2} I_0\left(2 r \textbf{r}_u\right) r d r\\
%&=e^{{-\textbf{r}_u}^2}\sum_n \frac{2 \textbf{r}_u^{2 n}}{(n!)^2} \int_0^{\infty} r^{2 n+1} e^{-r^2} d r=e^{{-\textbf{r}_u}^2}\sum_n \frac{\textbf{r}_u^{2 n}}{n!}= 1 \quad.
%\end{aligned}
%\end{equation}
%Here, $I_0$ is the modified Bessel function order 0. 
Consequently, the final result for the entropy is given by:
\begin{equation}
\begin{aligned}
S_L=1-\operatorname{tr}\left(\rho_\alpha^2\right)=\frac{1}{2}-\frac{1}{2} e^{-\left|\alpha_u-\alpha_d\right|^2}=\frac{1}{2}-\frac{1}{2} \exp\left(\left.-\int_0^{\infty} \left|\left.\frac{1-e^{-i \omega_k t}}{\omega_k} g \left(e^{-i \textbf{k} \cdot \textbf{r}_u}-e^{-i \textbf{k} \cdot \textbf{r}_d}\right)\right|\right.^2 d^3\textbf{k}\right)\right.,
\end{aligned}
\end{equation}

where, $g$ is the coupling constant of atoms; $g = g_\xi$. Now, we should calculate the integral in the exponential:
\begin{equation}
\begin{aligned}
& I =\int_0^{\infty} \left|\left.\frac{1-e^{-i \omega_k t}}{\omega_k} g \left(e^{-i \textbf{k} \cdot \textbf{r}_u}-e^{-i \textbf{k} \cdot \textbf{r}_d}\right)\right|\right.^2d^3\textbf{k}=\int_0^{\infty} \frac{8 \pi G m^2}{c \hbar k^3}\left(\left.1-\cos \left(\omega_x t\right)\right)\left(1-\cos \left(\textbf{k}\cdot\left(\textbf{r}_u-\textbf{r}_d\right)\right)\right)\right.d^3 \textbf{k} \quad.
\end{aligned}
\end{equation}
\\
Before deriving the above integral, we should use an approximation by recognizing that the timescale of the interaction is sufficiently small, making
$\cos \left(\omega_k t\right)$ vary insignificantly over this interval. Therefore:

\begin{equation}
\begin{aligned}
    \left(1-\cos \left(\omega_k t\right)\right)=(1-\cos (c k t)) \approx 1 \quad.
\end{aligned}
\end{equation}
Putting this approximation and integrating over the angular part, we have:\\
\begin{equation}
I \approx\int_0^{\infty}\int_{-1}^{+1}\frac{8 \pi G m^2}{c \hbar} \left(1-\cos \left(kr\cos \theta\right)\right) 2 \pi d(\cos \theta) dk = \frac{32 \pi^2 G m^2}{c \hbar k}\int_0^{\infty} \frac{kr-\sin{(kr)}}{(kr)^2}d(kr)\quad,\\
\end{equation}
where, we defined $\textbf{r} = \textbf{r}_u - \textbf{r}_d$. Then, we define the wave number of the Plank energy as a natural cutoff for the integral:
\begin{equation}
    k_{Planck} = \frac{\hbar c}{E_{Planck}} \sim 10^{32} m^{-1},
\end{equation}
Therefor, the result of the integral and the entropy would be:
\begin{equation}
I \approx \frac{32 \pi^2 G m^2}{c \hbar} \int_0^{\Lambda} \frac{\left(x-\sin \left(x\right)\right)}{x^2} d x \approx \frac{10^3 \pi^2 G m^2}{c \hbar}\approx 10^4 \frac{m^2}{m_p^2} \quad. \\
\end{equation}
Putting the atomic mass of the Rubidium atom we have:\\
\begin{equation}
\begin{aligned}
& I \approx 10^4 \times \frac{\left(16 \times 10^{-27} kg\right)^2}{\left(2.2 \times 10^{-8} kg\right)^2} \approx 10^{-29} \Rightarrow S_L \sim 10^{-29} \quad.
\end{aligned}
\end{equation}

\end{document}